\documentclass[%
 reprint,
 amsmath,amssymb,
 aps,
]{revtex4-2}

\usepackage{graphicx}
\usepackage{dcolumn}
\usepackage{bm}
\usepackage{mathptmx}
\usepackage{hyperref}
\usepackage{xcolor}
\usepackage{bbold}
\definecolor{darkblue}{rgb}{0, 0.3, 0.4}
\hypersetup{
    colorlinks=true,
    linkcolor=darkblue,
    filecolor=darkblue,      
    urlcolor=darkblue,
    citecolor = darkblue,
    pdftitle={Einstein gravity as a theory with a SL(2,C) connection double copy},
    pdfpagemode=FullScreen,
    }

\newcommand{\hookuparrow}{\mathrel{\rotatebox[origin=c]{90}{$\hookrightarrow$}}}
\newcommand{\hookdownarrow}{\mathrel{\rotatebox[origin=c]{-90}{$\hookrightarrow$}}}

\newcommand{\specialcell}[2][c]{%
  \begin{tabular}[#1]{@{}c@{}}#2\end{tabular}}
\renewcommand{\d}{\mathrm{d}}

\begin{document}

\preprint{APS/123-QED}

\title{Einstein gravity as a theory with a $SL(2,\mathbb{C})$ connection double copy}

\author{Vojtěch Witzany}
\affiliation{%
 School of Mathematics and Statistics, University College Dublin, Belfield, Dublin 4, Ireland\\
}%

\date{\today}

\begin{abstract}
Results ranging from Ashtekar variables to the perturbative Bern-Carrasco-Johansson (BCJ) double copy suggest a deep relation between Yang-Mills theory and Einstein gravity. I examine this relation by writing down the tetradic Palatini action for Einstein gravity and covariantly decomposing its variables into two $SL(2,\mathbb{C})$ connections and two soldering forms. This leads to a conjecture that the BCJ double copy can be understood through a ``holomorphic trail'', a series of injection maps between two copies of $SU(2)$ theories and Einstein gravity. 
\end{abstract}

\maketitle


\section{\label{sec:level1}Introduction}

For more than a century now, Einstein gravity has been passing more and more stringent tests on Earth, in the Solar system, and beyond  \cite{Will:2014kxa,Touboul:2017grn,Creminelli:2017sry,LIGOScientific:2019fpa,Wex:2020ald}. Its equations were first formulated directly in tensorial form as a non-linear partial differential equation for the space-time metric $g_{\mu\nu}$ \cite{1915SPAW.......778E}. Shortly thereafter it was realized that it can be also formulated through the so-called Einstein-Hilbert action, which is proportional to an integral of the Ricci scalar of the metric over the invariant space-time volume \cite{Hilbert:1915tx}. This formulation has some drawbacks, such as the fact that the action is non-polynomial in $g_{\mu\nu}$. Amongst other things, this implies that in a naive perturbative expansion of Einstein gravity around a flat background, there arise new $n$-point vertices at every consecutive order in the corresponding Feynman diagrams.

\textit{First order formalism and tetradic Palatini.} Some of the issues with the Einstein-Hilbert action can be remedied by using a first-order approach due to Palatini, which views the connection and the metric as independent degrees of freedom \cite{palatini1919deduzione,Deser:1969wk}. Specifically, when the degrees of freedom are expressed in terms of local-frame fields and a spin connection, one obtains a tetradic Palatini action (see e.g. \cite{Ashtekar:1991hf,Ashtekar:2004eh,Corichi:2016zac}), which is at most quartic in the aforementioned variables. Importantly, the tetradic Palatini action can be understood as a starting point when transforming to the so-called Ashtekar-Barbero variables and Loop quantization of Einstein gravity \cite{Ashtekar:1986yd,BarberoG:1993xfb, Ashtekar:2004eh}. By adding the so-called Holst term to the tetradic Palatini action (which does not change the classical equations of motion), one can obtain the self-dual Palatini action which, under a 3+1 split, leads to a Hamiltonian field theory of a Yang-Mills type set of variables \cite{Samuel:1987td,jacobson1987left,holst1996barbero}. The issue of this procedure is that the correspondence to a Yang-Mills phase space is restricted to the Hamiltonian formalism and a manifest connection to a covariant Yang-Mills Lagrangian is unclear.

\textit{KLT relations.} A possible new twist to this story arose quite recently. It is not uncommon that classes of solutions of various non-linear field theories can be formally related to each other. For example, a certain class of axially symmetric and stationary solutions of Yang-Mills-Higgs theory can be shown to be equivalent to solutions of Einstein equations in stationary axisymmetric vacuum space-times \cite{Singleton:1995xc}. However, relations useful in the analysis of \textit{generic} dynamical processes within the theories are far more precious. An example of such relations was given by Kawai, Lewellen and Tye, who have shown universal relations (now known as KLT relations) between tree-level gravitational scattering amplitudes and gauge-theory amplitudes \cite{Kawai:1985xq}. In return, this can be used to generate higher-loop diagrams for quantum gravities without reference to a Lagrangian via the so-called unitarity methods (see Ref. \cite{Bern:2002kj} for details and references). The general pattern in such constructions is that one uses two (possibly identical) Yang-Mills theories to generate the gravitational amplitude.

\textit{BCJ double copy.} The KLT relations have been further enhanced by a conjecture that every Yang-Mills theory can be put in a representation such that the kinematic numerators of its amplitudes satisfy a certain set of Jacobi-like identities similar to those of the color factors \cite{Bern:2008qj,Bern:2010ue,Bern:2010yg}. When put in this form, one can consequently take well-defined ``products'' of sets of two Yang-Mills amplitudes to obtain amplitudes in gravity theories. This is the Bern-Carrasco-Johanssen (BCJ) double-copy construction. Nevertheless, the composing pure Yang-Mills theories have ``too many'' degrees of freedom to represent a pure gravity, and additional massless fields such as an axion and a dilaton typically crop up in the resulting double copy. Fascinatingly enough, this can be resolved by introducing sources into the Yang-Mills theory and treating them as ghosts in the double copy \cite{Johansson:2014zca,Johansson:2019dnu}. The sum of the aforementioned methods has recently been used in a \textit{tour de force} set of computations of scattering of massive-particles in Einstein gravity up to 4th post-Minkowskian order \cite{Bern:2019nnu,Bern:2019crd,Bern:2021dqo}.

\section{Idea and summary of paper}

The successes but also the seemingly arbitrary structure of the BCJ double copy lead to the natural question: Is there a sense in which the BCJ double-copy could be derived ``from the top" instead of being an ad hoc procedure on the level of amplitudes? Or more specifically: Could one write a Lagrangian generating the Einstein equations in a form in which the double copy is manifest? I take some steps towards answering this question in this Letter (see also the works of \citet{Cheung:2016say,Cheung:2017kzx} for other approaches to the question).

It is now well know that the complexified Lie algebra of the Lorentz group $\mathfrak{so}(3,1)_{\mathbb{C}}$ can be written as a direct sum of two copies of the complexified Lie algebra of the rotation group $SU(2)$,  $\mathfrak{so}(3,1)_{\mathbb{C}} \simeq \mathfrak{su}(2)_\mathbb{C} \oplus \mathfrak{su}(2)_\mathbb{C}$. This leads to the labelling of finite-dimensional irreducible representations of the Lorentz group by the highest weights of two irreducible $SU(2)$ representations as $(m,n)$  (see, e.g., Refs. \cite{sexl2012relativity,carmeli2001classical}). 

A tetrad \textit{vector} basis $e^A_\mu$ and the dynamics of the related spin connection can be formally viewed as the local gauge theory of the $(1/2,1/2)$ (vector) representation of the Lorentz group. One could try and repeat this procedure by using a \textit{reducible} representation such as the $(1/2,0)\oplus(0,1/2)$ of the Lorentz group by using a tetrad \textit{Dirac-spinor} basis. One would then expect a natural decomposition of the dynamics into the two irreducible pieces of the representation. As relegated completely to the Supplemental material, this indeed turns out to be the case, but requires space-time metrics of signature $(--++)$. 

Consequently, I instead present in Section \ref{sec:decomp} of this Letter a procedure that leads to an equivalent result and can be well realised within $(-+++)$ signature space-times. I start with a tetrad formalism and decompose only the \textit{spin connection} related to the tetrad, which transforms in the $(1,0)\oplus(0,1)$ representation of the Lorentz group. As a result, a double copy of $SL(2,\mathbb{C})$ connections is also recovered in a fully covariant manner, and this is manifest in the resulting action.

As I then discuss in Section~\ref{sec:discuss}, this has intriguing connections both to the BCJ double copy and to Ashtekar-Barbero variables. In particular, I conjecture that the BCJ double copy can be understood through a ``holomorphic trail" between relativity and two copies of $SU(2)$ gauge theories as outlined in Table~\ref{tab:holotwist}.

\section{Decomposition of tetradic Palatini action} \label{sec:decomp}
{\em Notation and conventions:} Unless specified otherwise, the signature of the space-time metric $g_{\mu\nu}$ is $(-+++)$, and geometric units $G=c=1$ are used throughout. $A,B,C,D=0,...,3$ are tetrad indices belonging to the internal Lorentz representation, $\mu,\nu,\kappa,\lambda = 0,...,3$ are space-time indices, and $I,J,K,L=1,2,3$ are ``spatial" or $SL(2,\mathbb{C})$ internal indices. I use square brackets around indices to denote antisymmetrization, and round brackets to denote symmetrization.

\subsection{Tetrad basis Palatini action}

Consider a frame $e_\mu^A$ such that it satisfies the completeness relation
\begin{align}
    e^A_\mu e^B_\nu g^{\mu\nu} = \eta^{AB}\,,
\end{align}
where $\eta^{AB} = \rm{diag}(-1,1,1,1)$ is the Minkowski tensor. Then it is easy to show that 
\begin{align}
    e^A_\mu e^B_\nu \eta_{AB} = g_{\mu\nu}\,. 
\end{align}
In fact, in the tetrad formalism the tetrad is viewed as the primary object and the metric $g_{\mu\nu}$ and all the related objects as generated from it by the equation above. We can also define the covariant basis $e^\nu_B$ as the matrix inverse of $e_\mu^A$, that is $e^\nu_B e^B_\mu = \delta^\mu_\nu$, $e^\nu_A e^B_\nu = \delta^A_B$. Now I define the usual spin connection $\omega_\mu{}^{A}{}_B = e^A_{\kappa ; \mu} e_{B}^\kappa = -e^A_{\kappa} e_{B}^\kappa{}_{; \mu}$, where the semi-colon corresponds to the tensorial Levi-Civita derivative defined by $g_{\mu\nu;\kappa} = 0$. One can raise  or lower the index on the connection by $\eta^{BC}$ to obtain $\omega_\mu{}^{AB} = -\omega_\mu{}^{BA}$. Now the curvature associated to the connection is defined as
\begin{align}
    \Omega_{\mu\nu}{}^{AB} = \partial_\mu \omega_\nu{}^{AB} - \partial_\nu \omega_\nu{}^{AB} + \omega_\nu{}^{AC} \omega_\mu{}_C{}^{B} - \omega_\mu{}^{AC} \omega_\nu{}_C{}^{B} \,.
\end{align}
The Riemann tensor associated with the Levi-Civita connection acting only on tensors is recovered as
\begin{align}
    R_{\mu\nu\kappa\lambda} =  \Omega_{\mu\nu AB} e^A_\kappa e^B_\lambda \,,
\end{align}
and contractions can be formed likewise.

Now consider the tetradic Palatini action formulated in terms of frame forms $\boldsymbol{e}^A \equiv e^A_\mu \d x^\mu$:
\begin{align}
    S_{\rm TP} = \int \epsilon_{ABCD} \boldsymbol{e}^A \wedge \boldsymbol{e}^B \wedge \boldsymbol{F}^{CD}\,,
\end{align}
where $\epsilon_{ABCD}$ is the permutation symbol and the curvature form $\boldsymbol{F}^{AB} = F_{\mu\nu}{}^{AB} \d x^\mu \wedge \d x^\nu/2$ corresponds to an auxiliary connection $\kappa_\mu{}^{AB}$ as
\begin{align}
     F_{\mu\nu}{}^{AB} =\partial_\mu \kappa_\nu{}^{AB} - \partial_\nu \kappa_\nu{}^{AB} + \kappa_\nu{}^{AC} \kappa_\mu{}_C{}^{B} - \kappa_\mu{}^{AC} \kappa_\nu{}_C{}^{B}\,.
\end{align}
It is well known (see e.g. \cite{Ashtekar:1991hf,Ashtekar:2004eh,Corichi:2016zac}) that the variation of the action $S_{\rm TP}$ with respect to $\kappa$ yields that the auxiliary connection has to be the spin connection of the tetrad on-shell $\kappa_\nu{}^{AB} = \omega_\nu{}^{AB}$. Further variation with respect to $e^A_\mu$ then yields Einstein equations in the tetrad frame.

\subsection{Decomposition of connection}
Consider a general antisymmetric tensor $E^{AB} = -E^{BA}$ and its Lorentz transform
\begin{align}
    E'^{CD} = E^{AB} \Lambda^C{}_A \Lambda^D{}_B = E^{AB} \Lambda^{[C}{}_{[A} \Lambda^{D]}{}_{B]} \equiv E^{AB} L^{CD}{}_{AB}\,, 
\end{align}
where $\Lambda^C{}_A$ is the representation of the Lorentz transform in the vector representation and $L^{CD}{}_{AB}$ can be viewed as the representation of the Lorentz transformation on anti-symmetric tensors. One can define soldering symbols as
\begin{align}
    P^I_{\pm}{}_{AB} = \frac{1}{2}\left( \frac{1}{2} \epsilon_{IJK} \delta^J_{[A} \delta^K_{B]} \pm i \delta^0_{[A} \delta^I_{B]} \right)\,, \label{eq:solder} \\ 
    \tilde{P}^{AB}_{\pm}{}_I = \frac{1}{2} \epsilon_{IJK} \delta_J^{[A} \delta_K^{B]} \pm i \delta_0^{[A} \delta_I^{B]}\,. \label{eq:desolder}
\end{align}
Then one can decompose any antisymmetric tensor as 
\begin{align}
    E^I_{\pm} = E^{AB} P^I_{\pm}{}_{AB} \,,
\end{align}
and recover it from the $\pm$ pieces as
\begin{align}
    E^{AB} = \tilde{P}^{AB}_{+}{}_I E^I_- + \tilde{P}^{AB}_{-}{}_I E^I_+\,,
\end{align}
where one should note the alternating sign in the labels on the right-hand side. The index $I$ can be viewed as running only from 1 to 3, and the action of the Lorentz transform on the $E^I_{\pm}$ is
\begin{align}
\begin{split}
    & E'^M_\pm =  E'^{CD} P^M_{\pm}{}_{CD} = E^{AB} L^{CD}{}_{AB}  P^M_{\pm}{}_{CD} \\
    & = (\tilde{P}^{AB}_{+}{}_I E^I_- + \tilde{P}^{AB}_{-}{}_I E^I_+)L^{CD}{}_{AB}  P^M_{\pm}{}_{CD} \,.
\end{split}
\end{align}
The key insight is that for any Lorentz transform it holds that (see e.g. \cite{sexl2012relativity})
\begin{align}
    \tilde{P}^{AB}_{+}{}_I L^{CD}{}_{AB}  P^M_{+}{}_{CD} = \tilde{P}^{AB}_{-}{}_I L^{CD}{}_{AB}  P^M_{-}{}_{CD} = 0\,.
\end{align}
As such, the 3-component objects $E^I_{\pm}$ transform as
\begin{align}
    & E'^M_\pm = E^I_\pm \lambda^M_\pm{}_I\,,\\
    & \lambda^M_\pm{}_I \equiv  \tilde{P}^{AB}_{\mp}{}_I L^{CD}{}_{AB}  P^M_{\pm}{}_{CD}\,.
\end{align}
Furthermore, the matrices $\lambda^M_\pm{}_I$ correspond to two inequivalent representations of $SL(2,\mathbb{C})$, specifically the adjoint representation and its conjugate. This is a restatement of the fact that antisymmetric tensors transform in the $(1,0)\oplus (0,1)$ representation of the Lorentz group. One last note is that for real antisymmetric tensors it always holds that
\begin{align}
    E^I_+ = (E^I_-)^\dagger \,.
\end{align}
It may seem that the decomposition is non-covariant in the sense that one can choose any Lorentz ``pre-boost" before applying it. However, the action of the (continuous) Lorentz group is faithfully represented within each $\pm$ copy, and the procedure of the decomposition and recovery of the original tensor commute with Lorentz transforms. In other words, one can first boost and then apply the decomposition, or first apply the decomposition and only then boost, and the result will be the same. In this sense, the decomposition actually \textit{is} fully covariant.  

Now let us decompose the auxiliary spin connection into plus and minus copies $A^I_\pm$ as
\begin{align}
    \kappa_\nu{}^{AB} = \sum_{\pm} \tilde{P}^{AB}_{\mp}{}_I A_{\pm \nu}^I \,.
\end{align}
Now the curvature can also be decomposed as 
\begin{align}
    F_{\mu\nu}^{AB} =  \sum_{\pm} \tilde{P}^{AB}_{\mp}{}_I F_{\pm \mu \nu}^I\,.
\end{align}
The key statement is that $F_{\pm \mu \nu}^I$ can actually be obtained from the individual spin-connection pieces $A^I_{\pm \nu}$:
\begin{align}
    F_{\pm \mu \nu}^I = \partial_\mu A^I_{\pm \nu} - \partial_\nu A^I_{\pm \mu} \pm i \epsilon_{IJK} A^J_{\pm \mu} A^K_{\pm \nu}\,.
\end{align}
In other words, the connections $A^I_{\pm \nu}$ can be understood as complexified $\mathfrak{su}(2)$ connections in the adjoint and conjugate adjoint representations ($\simeq \mathfrak{sl}(2,\mathbb{C})$), and the decomposed curvatures can be understood as gauge-invariant field-strengths of the individual gauge connections $A^I_{\pm \nu}$.

Finally, the tetradic Palatini action becomes
\begin{align}
    & S_{\rm TP} = \int \boldsymbol{\Pi}^+_I \wedge \boldsymbol{F}^{I}_+ + \boldsymbol{\Pi}^-_I \wedge \boldsymbol{F}^I_- \,, \label{eq:decac}\\
    & \boldsymbol{\Pi}^{\pm}_I \equiv  \epsilon_{ABCD} \tilde{P}^{CD}_\mp{}_{I} \boldsymbol{e}^A \wedge \boldsymbol{e}^B \,,\\
    & \boldsymbol{F}^{I}_\pm \equiv  \frac{1}{2} F_{\pm \mu \nu}^I \d x^\mu\wedge \d x^\nu\,.
\end{align}
Here the forms $\boldsymbol{\Pi}^\pm_I$ can be viewed as the true {\em dynamical} soldering forms that solder antisymmetric tensors $E^{\mu\nu}$ into the internal $(1,0)\oplus(0,1)$ Lorentz representation.
\section{Discussion} \label{sec:discuss}
\subsection{Relation to the BCJ double copy}
The action \eqref{eq:decac} is at most quartic in its dynamic variables, and refers to two $SL(2,\mathbb{C})$ connections $A_{\pm}$ in the (conjugate) adjoint representation (I drop indices in this section for clarity of discussion). As such, it is a prime candidate for the understanding of the emergence of the BCJ double copy. Specifically, I propose here how the complex extension of the action \eqref{eq:decac} is the most likely candidate to understand the BCJ double copy.

\begin{table*}
\begin{center}
\begin{tabular}{|c|}
\hline \\
\begin{tabular}{c c c}
     \specialcell{$SU(2,\mathbb{R})$ gauge th. living on $\mathbb{M}_{4,\mathbb{R}}$,\\ symmetry of spct. $SO(3,1,\mathbb{R})$} 
     & 
     $\times$ 
     & 
     \specialcell{$SU(2,\mathbb{R})$ gauge th. living on $\mathbb{M}_{4,\mathbb{R}}$,\\ symmetry of spct. $SO(3,1,\mathbb{R})$} 
     \vspace{0.5em} \\ 
     \specialcell{(gauge fields $A^a_\mu$ real) \\ $\hookdownarrow$}  
     & 
     & 
     \specialcell{(gauge fields $A^a_\mu$ real) \\ $\hookdownarrow$}  
     \vspace{0.5em} \\ 
     \specialcell{$SU(2,\mathbb{C})\simeq SL(2,\mathbb{C})$ gauge th. living on $\mathbb{M}_{4,\mathbb{R}}$, \\ symmetry of spct. $SO(3,1,\mathbb{R})$}
     & 
     $\times$ 
     & 
     \specialcell{$SU(2,\mathbb{C})\simeq SL(2,\mathbb{C})$ gauge th. living on $\mathbb{M}_{4,\mathbb{R}}$, \\ symmetry of spct. $SO(3,1,\mathbb{R})$} \vspace{-0.5em}
     \\ 
     \multicolumn{3}{c}{\specialcell{$\underbrace{\hspace{37em}}$ \\ (Action of each $SO(3,1,\mathbb{R})$ mapped to half of spct. symmetries of target) \\ $\hookdownarrow$} } \vspace{0.25em}
     \\
     \multicolumn{3}{c}{\specialcell{$SL(2,\mathbb{C})\times SL(2,\mathbb{C})$ gauge th. on asymptotically $\mathbb{M}_{4,\mathbb{C}}$, \\ (Asymptotic) spct. symmetry $SO(3,1,\mathbb{C})$, homeomorphic to $SO(3,1,\mathbb{R})\times SO(3,1,\mathbb{R})$}}
     \vspace{0.25em} \\
     \multicolumn{3}{c}{\specialcell{$\hookuparrow$ \\ (analytic continuation from real coordinates)}}
    \vspace{0.5em} \\
     \multicolumn{3}{c}{\specialcell{$SL(2,\mathbb{C})\times SL(2,\mathbb{C})$ gauge th. (complex-metric relativity) on asymptotically $\mathbb{M}_{4,\mathbb{R}}$, \\ (Asymptotic) spct. symmetry $SO(3,1,\mathbb{R})$}} \vspace{0.25em}\\
     \multicolumn{3}{c}{\specialcell{$\hookuparrow$ \\ (tetrad and metric are real, $(A_+)^\dagger = A_-$)}} \vspace{0.5em}\\
     \multicolumn{3}{c}{\specialcell{$SL(2,\mathbb{C})$ gauge th. (real-metric relativity) on asymptotically $\mathbb{M}_{4,\mathbb{R}}$, \\ (Asymptotic) spct. symmetry $SO(3,1,\mathbb{R})$}} \\
     \\
\end{tabular} 
\\
\hline
\end{tabular}
\end{center}
\caption{The diagram of the ``holomorphic trail'' along which information is exchanged between two theories with $SU(2)$ connections and real general relativity. The injection maps $\hookrightarrow$ are briefly summarized in the brackets. $\mathbb{M}_{4,\mathbb{R}/\mathbb{C}}$ stands for real and complexified Minkowski space-time. The condition that the $SU(2)$ gauge fields $A^a_\mu$ are real refers to convention where gauge-group generators $T^a$ are Hermitian.} \label{tab:holotwist}
\end{table*}

\textit{Complexification of metric.} We can understand the $SL(2,\mathbb{C})$ connections $A_{\pm} $ as independent real degrees of freedom and treat them as such in the variation. However, once varying the action \eqref{eq:decac} with respect to them (and assuming a real tetrad and metric), it becomes clear that they have to be complex and that
\begin{align}
    A_{+} = A_{-}^\dagger\,.
\end{align}
That is, the premise of the variation principle is violated by the solution, and one should instead write the action from the outset in the manifestly real form
\begin{align}
    & S_{\rm TP} = \int \boldsymbol{\Pi}^+_I \wedge \boldsymbol{F}^{I}_+ + (\boldsymbol{\Pi}^+_I \wedge \boldsymbol{F}^I_+)^\dagger \label{eq:redecac} \,. 
\end{align}
This form generates Einstein equations for real metrics. 

An alternative relevant to the BCJ double copy would be to understand the connections as independent and complex and vary the action \eqref{eq:decac} with respect to all of the four degrees of freedom $A_{+},A_{-},A^\dagger_{+},A_{-}^\dagger $. Note that the connections $A_\pm$ then directly correspond to the ``heavenly" and ``hellish" sectors of Einstein gravity in the complex space-time approach advocated by \citet{Plebanski:1977zz}. However, at the end of the day one wants to stay ``earthly" by restricting to physical degrees of freedom and separately imposing $A^I_{+} - (A^I_{-})^\dagger =0$ as a Lagrangian constraint. Upon quantization, this constraint can be enforced by the introduction of ghosts, very much how this was found to be necessary in the double copy procedure by \citet{Johansson:2014zca,Johansson:2019dnu}.  

\textit{Double Lorentz group.} There is, however, an even stronger hint that a full complex prolongation of relativity is likely the key to understanding the double copy as follows. Recall that the group $SL(2,\mathbb{C})$ is a double cover of the real Lorentz group $SO(3,1,\mathbb{R})$  \cite{sexl2012relativity,carmeli2001classical}. However, it also holds that the \textit{complexification} of the Lorentz group $SO(3,1,\mathbb{C})$ has a double cover $SL(2,\mathbb{C})\times SL(2,\mathbb{C})$ which is also a cover of $SO(3,1,\mathbb{R})\times SO(3,1,\mathbb{R})$. In other words, if we take the theory with a complexified metric and also allow for complex coordinates, asymptotically flat space-times enjoy a projective \textit{double Lorentz symmetry} that is naturally also presented in scattering amplitudes. This is important since,  as stressed by \citet{Cheung:2016say}, the double Lorentz symmetry is one of the key symmetries of the BCJ double copy. It should be noted, however, that relativity and pairs of theories with $SU(2)$ connections flow towards the same, fully complexified class of theories through a set of injection maps that are \textit{different} for each theory, as is detailed in the diagram in Table~\ref{tab:holotwist}. As such, there is a \textit{holomorphic trail} between relativity and double $SU(2)$ theories along which information is exchanged. Details of some of the less trivial injections and further discussion will be published in a separate paper. 

\textit{Open Issues.} Let me comment on a few more outstanding issues in understanding the full relation of this result to the BCJ double copy. It may seem that one should require the connections $A_{\pm}$ to be Yang-Mills connections in the sense that they have vanishing divergence or at least fulfill some analogous dynamics. This is not necessarily true, since the BCJ amplitude composition method actually requires the use of generalized gauge transforms that can be seen as deeper variable transforms, and the stripping of the color factors and replacing by kinematic factors in the procedure changes the dynamics on an even more fundamental level. A more direct understanding of the correspondence must then be established on the level of observables and amplitudes, which can likely be done only in very special gauges on either side of the correspondence. This task is well outside the scope of this Letter. 

A closely related open issue is also the counterpart of the soldering forms $\boldsymbol{\Pi}^I_\pm$ in the double copy. In a geometric sense, each of the $\pm$ sets represents a normalized complex triad of oriented 2D space-time surfaces. By examining the equations of motion, they can be seen as generating super-potentials for the connections. However, in the context of the 3+1 formalism tetrad-related objects quite surprisingly tend to instead play the role of canonical momenta conjugate to the connections. 
The spinorial procedure given in the Supplemental material demonstrates that there is a number of ways in which the soldering forms arise and that there is more than one route to obtain the action \eqref{eq:redecac}. This means that soldering forms do not need to be generated by tetrads at all and could instead be postulated directly with appropriate completeness relations in place. By comparing to the first-order formalism for Yang-Mills theory given by \citet{Deser:1969wk} (eq. (13a) therein), the soldering forms play the role similar to Deser's auxiliary curvature variable, but miss a ``kinematic'' term $\propto \Pi^I_{\mu\nu} \Pi^{\mu\nu}_I$ for this interpretation. Such terms could arise as topological terms necessary to differentiate the action., see Ref. \cite{Corichi:2016zac}.

\subsection{Relation to Ashtekar variables}

It is obvious that the decomposition into $A_{\pm}$ as appearing here can be viewed, in fact, as a fully covariant derivation of the Ashtekar connection without the need of adding a Holst term to the action as done in Refs \cite{Samuel:1987td,jacobson1987left,holst1996barbero}. To see that, consider that the self-dual connection $\mathcal{A}^{CD}$ used by Ashtekar is in our $(-+++)$ metric case (following approximately the notation of \cite{Ashtekar:2004eh} and dropping space-time indices)
\begin{align}
    \mathcal{A}^{CD} = \frac{1}{2}\omega^{AB}(\delta^{C}_{A}\delta^{D}_{B} + \frac{i}{2}\epsilon^{AB}{}_{CD})\,.
\end{align}
Then we have $A^{I}_+ = i \mathcal{A}^{0I}$. Also, one can compare equations (12) and (18) in ref. \cite{holst1996barbero}, and eq. \eqref{eq:solder} as given here to see that the straightforward correspondence with the 3-component version of the canonical Ashtekar variables. The key insight that makes this work different is that one does not need to use the entire self-dual connection and curvature and then restrict to its certain components in the 3+1 decomposition, as is done in the usual procedure \cite{Samuel:1987td,jacobson1987left,Ashtekar:2004eh}. Instead, one can refer directly to the covariant $(1,0)\oplus(0,1)$ decomposition already before the 3+1 split.

The main breakthrough presented by Ashtekar variables was the reduction of the Hamiltonian constraints in the 3+1 formalism into simple polynomial forms (see Ref. \cite{Rovelli:1997yv} for a historical overview). However, this came at the cost of complexifying the action and the metric, which had to be solved in an ad-hoc manner. This led, e.g., to the two-connection formalism of Barbero \cite{BarberoG:1993xfb,holst1996barbero}, where Barbero succeeded in a similar effort in the framework of a 3+1 \textit{real} formalism. It is unclear what is the relation of the herein presented formalism with that of Barbero. 

However, the fully covariant and real expression given in equation \eqref{eq:redecac} could circumvent these issues altogether, for instance, by quantizing in a path-integral approach. The fact that the KLT relation and the BCJ double copy emerged from the consideration of string scattering amplitudes means that further research into this question could also shed, at least in principle, some light on the long-sought link between String theory and Loop quantum gravity.

\begin{acknowledgments}
I would like to thank Justin Vines and Geoffrey Comp{\`e}re for feedback on the early manuscript. This work was supported by European Union’s Horizon 2020 research and innovation programme under grant agreement No 894881.
\end{acknowledgments}

\bibliography{apssamp}


\section*{Supplemental material: A Dirac-Palatini formulation of Einstein gravity}

Here I discuss a counter-part to the tetrad construction from the main paper by using a basis of Dirac spinors. 

\subsection{Dirac-spinor basis}

I will use the chiral representation of spinors and the Clifford algebra $\gamma^{(A} \gamma^{B)} = - \eta^{AB}$, in which the gamma matrices $\gamma^A$ satisfy
\begin{align}
\begin{split}
    & \gamma^0 = 
    \begin{pmatrix} 
    \mathbb{0}_2 & \mathbb{1}_2 \\
    \mathbb{1}_2 & \mathbb{0}_2
    \end{pmatrix} \,,\; \gamma^I = \begin{pmatrix} 
    \mathbb{0}_2 & \sigma^I \\
    \sigma^I & \mathbb{0}_2
    \end{pmatrix}\,, \\ & \sigma^I = \begin{pmatrix} 
    \delta_{I3} & \delta_{I1} - i \delta_{I2} \\
    \delta_{I1} + i \delta_{I2} & -\delta_{I3}
    \end{pmatrix}\,.
    \end{split}
\end{align}
where $\sigma^I$ are Pauli matrices. The choice of the chiral representation will be important because it makes the generators of the Lorentz group $S^{AB} = -S^{BA} = \gamma^{[A} \gamma^{B]}/2$ block-diagonal:
\begin{align}
    S^{0I} = 
    \begin{pmatrix}
    \sigma^I & \mathbb{0}_2 \\
    \mathbb{0}_2 &-\sigma^I \\
    \end{pmatrix}\,,\; 
    S^{IJ} = i \epsilon_{IJK} 
    \begin{pmatrix}
    \sigma^K & \mathbb{0}_2 \\
    \mathbb{0}_2 &\sigma^K \\
    \end{pmatrix}\,,
\end{align}
where $\epsilon_{IJK}$ is the permutation symbol. I call the part of the spinor upon which the upper block acts ``left" and the ones one which the lower ones act as ``right". That is, every upper-index spinor can be split into the form $\chi^a = (\chi_{\rm L}, \chi_{\rm R})$. The indices $a,b,c =1,...,4$ are spinor index transforming in the $(1/2,0)\oplus(0,1/2)$ spinor representation. Dirac-adjoint spinors are written with a lower index, $\bar{\chi}_b$, which transforms with the inverse Lorentz transform (that is, the dual representation).  

Consider now a basis of Dirac spinors $\psi_\mu^a$ in space-time such that they fulfill the completeness relations
\begin{subequations} \label{eq:complet}
\begin{align}
& (\bar{\psi}_\mu)_b \psi^a_\nu g^{\mu\nu} = \delta^a_b \,,
\label{eq:psidelt} \\
&\bar{\psi}_{[\mu} \psi_{\nu]} = 0 \,, \label{eq:psireal} 
\end{align}
\end{subequations}
where $(\bar{\chi})_c = (\chi^\dagger \gamma^0)_c = (\chi^a)^*\delta_{ab} (\gamma^0)^b{}_c $ is the Dirac adjoint. In this sense, the adjoint operation can be viewed as Lorentz-covariant index lowering and raising. (This requires representations where $\gamma^{0\dagger} = \gamma^0$.) 

The matrix $\psi_\mu^a$ has 16 complex components corresponding to 32 real degrees of freedom. The constraint \eqref{eq:psidelt} has 10 independent complex components (corresponding to 20 real constraints), and equation \eqref{eq:psireal} has 6 purely imaginary components (6 real constraints). As such, they leave only 6 free real parameters in $\psi_\mu^a$. Since all the constraints are Lorentz-invariant, the freedom in the non-trivial solutions to the constrains uniquely corresponds to the 6 parameters of the (internal) Lorentz group.  However, the equations \eqref{eq:complet} quickly yield that the completeness relations can only be satisfied in space-times with signature $(--++)$, since they require the existence of four linearly independent complex vectors $p_\mu,q_\mu,r_\mu,s_\mu$ such that $p^*_\mu p^\mu = q^*_\mu q^\mu = r^*_\mu r^\mu =s^*_\mu s^\mu =0$. 

Finally, the basis of Dirac spinors $\bar{\psi}_\mu^a$ can be viewed as generating the metric (with $(--++)$ signature) very much like a basis of tetrad vectors
\begin{align}
    \bar{\psi}_\mu \psi_\nu = g_{\mu\nu}\,. \label{eq:psitog}
\end{align}
 In the rest of this material I will treat the Dirac-spinor basis as the primary object, similarly to the tetrad-vector approach. The metric $g_{\mu\nu}$, and thus implicitly also its inverse $g^{\mu\nu}$, should then be understood as defined by the spinor frame via \eqref{eq:psitog}.

\subsection{The spinor connection}
Let me now define an affine spinor connection $\omega^a{}_{c\mu}$ by postulating a spinor-tensor covariant derivative $D_\mu$ such that mixed-index objects satisfy
\begin{align}
\begin{split}
    D_\mu \chi_\nu^{a...}{}_{b...} = & \nabla_\mu \chi_\nu^a + \omega^a{}_{c \mu} \chi_\nu^{c...}{}_{b...} + ...  \\ & - \omega^c{}_{b \mu} \chi_\nu^{a....}{}_{c....} - ... \,,
    \end{split}
\end{align}
where $\nabla_\mu$ is the purely tensorial Levi-Civita covariant derivative $\nabla_\mu g_{\nu\kappa} = 0$ that ignores the spinor indices. The relationship to the spinor frame is then deduced by further requiring that $D_\mu \psi_\nu^a = 0$, which yields 
\begin{align}
\omega^a{}_{b \mu}  = -(\bar{\psi}^\nu)_b \nabla_\mu \psi_\nu^a = \psi_\nu^a \nabla_\mu(\bar{\psi}^\nu)_b  \,,
\end{align}
where $\bar{\psi}^\nu = g^{\nu\kappa}\bar{\psi}_\kappa$. 

 The spinor curvature is defined by $((D_\mu D_\nu - D_\nu D_\mu)\chi)^a = (\Omega_{\mu\nu} \chi)^a$, which yields explicitly
\begin{align}
    \Omega^a{}_{c\mu\nu} = \partial_\mu \omega^a{}_{c \nu} - \partial_\nu \omega^a{}_{c \mu} + \omega^a{}_{b \nu} \omega^b{}_{c\mu} - \omega^a{}_{b\mu} \omega^b{}_{c\nu}\,, \label{eq:curvdef}
\end{align}
where reference to the Levi-Civita connection $\nabla_\mu$ vanishes due to $\nabla_\mu A_\nu - \nabla_\nu A_\mu = \partial_\mu A_\nu - \partial_\nu A_\mu$.

Thanks to the use of the chiral representation and the completeness relations \eqref{eq:complet}, the connection can only generate infinitesimal Lorentz transforms of block-diagonal form acting separately on the right and the left parts of the Dirac-spinor basis. The same statement is obviously true for the curvature. In other words, the connection and curvature will always be of the form
\begin{align}
    \omega^a{}_{b\mu} =
    \begin{pmatrix}
    \omega_{\rm L \mu} & \mathbb{0}_2 \\
    \mathbb{0}_2 & \omega_{\rm R \mu} 
    \end{pmatrix}\,, \;
    \Omega^a{}_{b \mu\nu} = 
    \begin{pmatrix}
    \Omega_{\rm L \mu\nu} & \mathbb{0}_2 \\
    \mathbb{0}_2 & \Omega_{\rm R \mu \nu} \label{eq:blockdiag}
    \end{pmatrix}\,,
\end{align}
where $\Omega_{\rm L/R}$ is given in terms of $\omega_{\rm L/R}$ in an identical manner as in \eqref{eq:curvdef} only with spinor indices running over more restricted values (in particular, the commutator $[\omega_\nu,\omega_\mu]$ does not mix between the blocks).

The relationship between the curvature of the spinor connection and that of the tensorial Levi-Civita connection is established by direct computation as
\begin{align}
    \Omega^a{}_{b \mu\nu} = R_{\mu\nu}{}^{\kappa \lambda} \psi^a_\kappa (\bar{\psi}_\lambda)_b\,.
\end{align}
In particular, the Ricci tensor and scalar of $\nabla_\mu$ are obtained as
\begin{align}
    & R_{\mu\nu} = \bar{\psi}^\mu_a\Omega^a{}_{\mu\nu} \psi^b_\lambda \label{eq:riccit}\,,\\
    & R = {\bar{\psi}^\mu}_a \Omega^a{}_{b \mu\nu} \psi^{\nu b}\,. \label{eq:riccis}
\end{align} 

\subsection{Dirac-Palatini action}

Consider the action
\begin{align}
    S_{\rm DP}[\psi_\mu^a,\kappa^a{}_{b\mu}] = \int \Gamma {\bar{\psi}^\mu}_a F^a{}_{b \mu\nu} \psi^{\nu b} \d^4 x\,, \label{eq:SDP}
\end{align}
where $\Gamma = \sqrt{-{\rm det}(\bar{\psi}_\mu \psi_\nu)}$ and
$F^a{}_{b \mu \nu}$ is the curvature of an auxiliary spin connection $\kappa^a{}_{b\mu}$:
\begin{align}
    F^a{}_{b\mu\nu} = \partial_\mu \kappa^a{}_{b \nu} - \partial_\nu \kappa^a{}_{b \mu} + \kappa^a{}_{c \nu} \kappa^c{}_{b\mu} - \kappa^a{}_{c\mu} \kappa^c{}_{b\nu}\,. \label{eq:auxcurv}
\end{align}
It is assumed that the variation of $\psi_\mu^a$ is carried out while respecting the completeness relations \eqref{eq:psidelt} and \eqref{eq:psireal}, and that the connection induces infinitesimal Lorentz transforms on any spinor, so that it is in the general block-diagonal form \eqref{eq:blockdiag}.

The action \eqref{eq:SDP} refers to the metric determinant and to the metric inverse and is thus non-polynomial in $\psi^a_\mu$. Even though it is in principle possible to reformulate the action $S_{\rm{DP}}$ using differential forms so that it is purely polynomial in the variables $\psi^a_\mu$ and $\kappa^a{}_{b\nu}$, it seems that this also requires violating manifest Lorentz covariance of the expressions. Hence, I will only use the non-polynomial form given in equation \eqref{eq:SDP} in this Supplemental material.

The statement to prove now is that the action \eqref{eq:SDP} generates equations equivalent to vacuum Einstein equations under the requirement of vanishing variation with respect to $\psi^a_\mu$ and $\kappa^a{}_{b\nu}$. 

Variation with respect to $\kappa^a{}_{b\nu}$ yields
\begin{align}
    (\bar{\psi}^\mu)_a \psi^{\nu b} (\delta^a_d  \kappa^e{}_{b [\mu} \delta^\lambda_{\nu]} + \delta^e_b  \kappa^a{}_{d [\nu} \delta^\lambda_{\mu]}) = \nabla_\gamma ((\bar{\psi}^{[\gamma})_d \psi^{\lambda] e})\,, \label{eq:kappavar}
\end{align}
It is easy to verify that this equation has a particular solution given by the spinor connection $\kappa^a{}_{b\nu} =\omega^a{}_{b \nu}$. Then one can search for homogeneous solutions $C^a{}_{b\nu} = \kappa^a{}_{b\nu} - \omega^a{}_{b \nu}$ which satisfy equation \eqref{eq:kappavar} with a zero right-hand side. I have verified by using a brute-force calculation in a specific frame that spinor frames fulfilling the completeness relations \eqref{eq:complet} allow no homogeneous solutions for equation \eqref{eq:kappavar}. In other words, the variation with respect to $\kappa^a{}_{b\mu}$ yields uniquely $\kappa^a{}_{b\mu} = \omega^a{}_{b \mu}$. By substituting this result into $S_{\rm DP}$ and considering equation \eqref{eq:riccis} one sees that it becomes proportional to the usual Einstein-Hilbert action.

Now, to vary $S_{\rm DP}$ with respect to $\psi^a_\mu$ we need to apply variations that do not violate completeness relations. Since these degrees of freedom are only variations to Lorentz transforms of the spinors (with respect to which the action is invariant), and variations that uniquely map to metric variations, the procedure straightforwardly yields the vacuum Einstein equations
\begin{align}
    R_{\mu\nu} - \frac{1}{2} R g_{\mu\nu} = 0 \,,
\end{align}
where the Ricci tensor and scalar are given in equations \eqref{eq:riccis} and \eqref{eq:riccit}.

\subsection{Left-right split of Dirac-Palatini action}

It is obvious by inspecting the block-diagonal form of the connection and the curvature \eqref{eq:blockdiag}, which, by assumption translates also to the auxiliary connection and curvature, that one can write the Dirac-Palatini action in the form:
\begin{align}
    S_{\rm DP}[\psi_\mu^a,\kappa^a{}_{b\mu}] = \int \Gamma (\chi^{*\mu}_{{\rm L}p} F^{p}_R{}_{q \mu\nu} \chi^{\nu q}_{\rm R} + \chi^{*\mu}_{{\rm R} p} F^p_L{}_{{q} \mu\nu} \chi^{\nu {q}}_{\rm L}) \d^4 x\,,
\end{align}
where I have introduced indices $p,q,p,q=1,2$ that transform in the $SL(2,\rm{C})$ transformation. Some of the indices transform in the conjugate representations, and the lower index position denotes that it transforms by inverse transforms (in the dual representation). 

Similarly to the tetrad approach, one can solder the connection and curvature into the adjoint representation, this time by Pauli matrices:
\begin{align}
    A^I_{\rm L/R} = (\sigma^I)^q{}_p \omega_{\rm L/R}^p{}_q\,,\;\; \omega_{\rm L/R}^p{}_q = \frac{1}{2}(\sigma^I)^p{}_q A^I_{\rm L/R}\,. \label{eq:spsolder}
\end{align}
It can be then shown from the fact that the connection induces infinitesimal Lorentz transforms that on-shell one necessarily has
\begin{align}
    (A^I_{\rm L})^* = -A^I_{\rm R}\,.
\end{align}
Independently, it can be  seen that curvatures obtained by the same projections fulfill
\begin{align}
    F^I_{}{}_{\mu\nu} = \partial_\mu A^I{}_\nu - \partial_\nu A^I{}_\mu + i\epsilon_{IJK} A^J_\mu A^K_\nu\,,
\end{align}
where the formula applies both to the left and the right curvature. Finally, one can rewrite the action as
\begin{align}
    & S_{\rm DP}[\psi_\mu^a,\kappa^a{}_{b\mu}] = \int \Gamma (F^I_{\rm L}{}_{\mu\nu} \Pi^{\mu\nu}_{\rm L}{}_I + F^I_{\rm R}{}_{\mu\nu} \Pi^{\mu\nu}_{\rm R}{}_I ) \d^4 x\,,\\
    & \Pi^{\mu\nu}_{\rm L}{}_I =  \chi^{*[\mu}_{{\rm L}p} \chi^{\nu] q}_{\rm R} (\sigma^I)^p{}_q\,,\\
    & \Pi^{\mu\nu}_{\rm R}{}_I = \chi^{*[\mu}_{{\rm R}p} \chi^{\nu] q}_{\rm L} (\sigma^I)^p{}_q \,.
\end{align}
Interestingly, $(\Pi^{\mu\nu}_{\rm L}{}_I)^* = -\Pi^{\mu\nu}_{\rm R}{}_I$. The fact that the objects are anti-conjugate follows from some choices of convention made during this derivation, specifically the choice to keep the soldering prescriptions in equation \eqref{eq:spsolder} identical for the left and right sectors. 

In summary, the Dirac-spinorial approach generates, at the cost of a $(--++)$ signature metric, an action completely analogous to the decomposed action \eqref{eq:redecac}.

\end{document}